\documentclass[prb,aps,twocolumn]{revtex4}
\usepackage{graphicx,amssymb,amsmath,color,psfrag}
\usepackage{amsthm}
\usepackage{amsfonts}
\usepackage{algorithmic}
\usepackage{enumerate}
\usepackage{latexsym}

\begin{document}
\newcommand{\dt}{\Delta\tau}
\newcommand{\al}{\alpha}
\newcommand{\ep}{\varepsilon}
\newcommand{\ave}[1]{\langle #1\rangle}
\newcommand{\have}[1]{\langle #1\rangle_{\{s\}}}
\newcommand{\bave}[1]{\big\langle #1\big\rangle}
\newcommand{\Bave}[1]{\Big\langle #1\Big\rangle}
\newcommand{\dave}[1]{\langle\langle #1\rangle\rangle}
\newcommand{\bigdave}[1]{\big\langle\big\langle #1\big\rangle\big\rangle}
\newcommand{\Bigdave}[1]{\Big\langle\Big\langle #1\Big\rangle\Big\rangle}
\newcommand{\braket}[2]{\langle #1|#2\rangle}
\newcommand{\up}{\uparrow}
\newcommand{\dn}{\downarrow}
\newcommand{\bb}{\mathsf{B}}
\newcommand{\ctr}{{\text{\Large${\mathcal T}r$}}}
\newcommand{\sctr}{{\mathcal{T}}\!r \,}
\newcommand{\btr}{\underset{\{s\}}{\text{\Large\rm Tr}}}
\newcommand{\lvec}[1]{\mathbf{#1}}
\newcommand{\gt}{\tilde{g}}
\newcommand{\ggt}{\tilde{G}}
\newcommand{\jpsj}{J.\ Phys.\ Soc.\ Japan\ }

\title{Magnetic impurity in the vicinity of a vacancy in bilayer graphene }
\author{J. H. Sun$,^{1}$ F. M. Hu$,^1$ H. K. Tang$,^{2}$ and H. Q. Lin$^1$}
\affiliation{$^1$Department of Physics and ITP, The Chinese University of Hong Kong, Hong Kong, China\\
$^2$COMP/Department of Applied Physics, Aalto University School of Science, P.O. Box 11000, FI-00076 Aalto, Espoo, Finland}

\begin{abstract}
We use quantum Monte Carlo method to study a magnetic impurity located next to a vacancy in bilayer graphene with Bernal stacking.
Due to the broken symmetry between two sublattices in bilayer system, there exist two different types of vacancy induced localized state.
We find that the magnetic property of the adatom located on the adjacent site of the vacancy depends on whether the vacancy belongs to A or B sublattice.
In general, local moment is more strongly suppressed if the vacancy belongs to the sublattice A when $\mu \sim 0$. We switch the values of the chemical potential and study the basic thermodynamic quantities and the correlation functions between the magnetic adatom and the carbon sites.

\end{abstract}
\date{\today}
\maketitle
\section{Introduction}
Interest in the defects induced localized states in monolayer and multilayer graphene has been greatly expanding \cite{Castro10,Pereira06, Ugeda10} driven by the enthusiasm to seek for ferromagnetism in graphene \cite{Harigaya01, Lehtinen04,Vozmediano05}.
Experimentally, one can introduce vacancies in carbon-based systems by irradiation to modify their properties \cite{Esquinazi03, Ohldag07, Krasheninnikov07, Gomez05}.
In monolayer graphene, vacancy leads to the formation of zero-energy quasi-localized states \cite{Pereira06} such that the local density of states (LDOS) of the neighboring sites would be greatly modified. Consequently, the behavior of a magnetic impurity in the vicinity of a vacancy shall be different either from that in a pure graphene or from that in a normal metal. Study on the consequences of an adjacent vacancy-adatom \cite{hu2012} showed that the local moment formation is strongly suppressed. However, in pure graphene where the density of states(DOS) vanishes at Dirac point, the Kondo effect does not occur in general\cite{Withoff90}, so a well-developed local moment will show up in undoped case \cite{Hu11}.

In Bernal stacked bilayer graphene (BLG), due to the interlayer hopping energies, the two sublattices are no longer equivalent. It has been reported that whether the defect is generated on A or B sublattice would greatly alters the defect induced magnetism and quasi-localized states \cite{Castro10, Ugeda10}.

In this paper, we study the magnetic properties of an Anderson impurity on the adjacent site of a vacancy in Bernal stacked bilayer graphene using the quantum Monte Carlo method based on the Hirsch-Fye algorithm \cite{Hirsch86}.

We invoke the Anderson impurity model, and the total Hamiltonian can be written as
\begin{equation}
\begin{aligned}
H=H_{0}+H_{1}+H_2+H_{vac}.
\end{aligned}
\end{equation}
$H_0$, $H_1$, $H_2$ are the tight-binding Hamiltonian of BLG, the impurity Hamiltonian and the hybridization between impurity and the carbon electrons, respectively. They can be written as
\begin{equation}
\begin{aligned}
H_0&=(-t\sum\limits_{<i,j>m}a_{mi}^{\dagger}b_{mj}-t_1\sum\limits_{j}a_{1j}^{\dagger}a_{2j})+H.c., \\
H_1&=\mu\sum\limits_{im}(a_{mi}^{\dagger}a_{mi}+b_{mi}^{\dagger}b_{mi}), \\
H_2&=(\varepsilon_d-\mu) d^{\dagger}d+V(c_{md}^{\dagger}d+H.c.),
\end{aligned}
\end{equation}
where $a_{m i\sigma}$($b_{m i\sigma}$) annihilates an electron with spin $\sigma$ at the site $R_{mia}$ ($R_{mib}$) on sublattice A(B) of graphene's hexagonal structure in the m-th layer. we use $t\approx2.8eV$ \cite{Rmp} as the energy unit. For the inter-layer hopping energies, according to the Slonczewski-Weiss-McClure parametrization \cite{Brandt88,Dresselhaus02}, $t_1\approx0.4eV$. $d_{\sigma}$ annihilates an electron with spin $\sigma$ at the impurity orbit. If the magnetic adatom is added on the top of sublattice A, $c_{md\sigma}=a_{md\sigma}$, otherwise $c_{md\sigma}=b_{md\sigma}$, where d denotes the location of the impurity adatom.
Without loss of generality, we assume the vacancy is located on layer 1. If the vacancy is adsorbed on sublattice A, the Hamiltonian $H_{Vac}$ induced by the vacancy can be written as
\begin{equation}
\begin{aligned}
H_{Vac}=&t\sum\limits_{j}(a_{10}^{\dagger}b_{1j}+H.c.)+t_1(a_{10}^{\dagger}a_{20}+H.c.)\\
&+\mu a_{10}^{\dagger}a_{10}.
\end{aligned}
\end{equation}
We would not consider about the effect of reconstruction of the remaining structure in the presence of a vacancy, since the properties of localized states are rather insensitive to the deformation \cite{Choi08}.

\section{Results}

We mainly test two different groups, namely, vacancy belongs to sublattices A and B, and the magnetic adatom is located on the adjacent site of the vacancy.
In all the results we show below, we apply the minimal model and use $t_1=0.2t$ and ignore other interlayer hopping energies since they are much smaller than $t_1$ and have minor effect on the behavior of magnetic adatom.


\begin{figure}[t]
\begin{center}
\includegraphics[scale=0.4, bb=200 20 400 510]{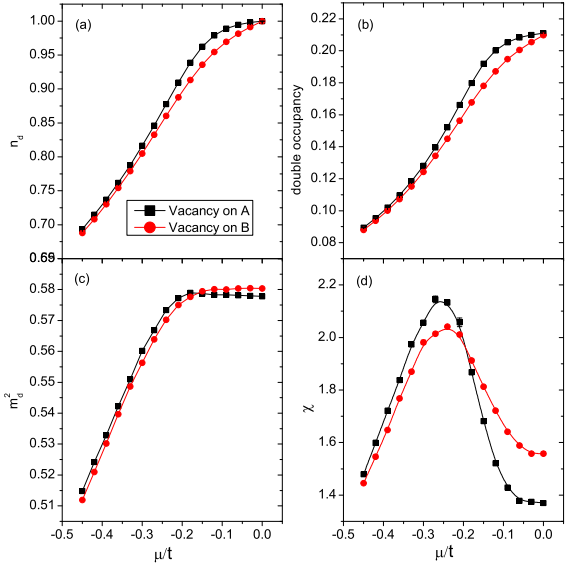}
\end{center}
\caption{(Color online). The various thermodynamic quantities for the two cases that vacancy embedded in sublattices A and B. The impurity is located on the nearest site of the vacancy on layer 1. (a) $n_{d}$ as a function of $\mu$. (b) $n_{d\uparrow}n_{d\downarrow}$ as a function of $\mu$. (c) $m_{d}^{2}$ as a function of $\mu$. (d) $\chi$ as a function of $\mu$. In all the plates we use $V=1.0t$, $U=0.8t$, $\varepsilon_{d}=-U/2$ and $\beta=1/T=40t^{-1}$.} \label{Fig:vacR0Layer1basicr}
\end{figure}

Shown in Fig. \ref{Fig:vacR0Layer1basicr} are the basic thermodynamic quantities of the impurity on the nearest neighbor of the vacancy for both the cases that the vacancy belongs to sublattices A and B. Shown in Fig. \ref{Fig:vacR0Layer1basicr} (a) are the values of charge $n_d$, and we can see that the A vacancy case shows slightly larger occupancy than B vacancy case near the zero-energy. The values of double occupancy $n_{d\uparrow}n_{d\downarrow}$ shown in Fig. \ref{Fig:vacR0Layer1basicr}(b) has the same order as that of $n_d$. We can see that as we lower the values of chemical potential from zero-energy, the values of $n_d$ and $n_{d\uparrow}n_{d\downarrow}$ are gradually decreased. However, the local moment squared shown in Fig. \ref{Fig:vacR0Layer1basicr}(c) is first increased and then decreased for the A vacancy case, while mostly unchanged for B vacancy case around the zero-energy. According to the fact that $n_{d\sigma}$ can either be zero or one, we have
\begin{equation}
m_d^2=n_d-2n_{d\uparrow}n_{d\downarrow} .
\end{equation}
The suppression of the local moment formation is mainly caused by the larger values of double occupancy $n_{d\uparrow}n_{d\downarrow}$ compared with the pure BLG case. We can also find that the impurity magnet moment for the B vacancy case is slightly larger than that of A vacancy case. As the chemical potential is lowered, they cross over each other and finally the difference between these two values vanishes. The difference is more obvious in the values of spin susceptibility as shown in Fig. \ref{Fig:vacR0Layer1basicr}(d). In principle, spin susceptibility depends not only on the local moment itself but also on the spin correlation with the conduction band electrons. The A vacancy leads to sharper resonance on the LDOS of the nearest neighbor around $E=0$, which will cause stronger screening of the local moment. From Fig. \ref{Fig:vacR0Layer1basicr}(d), we can also find that the spin susceptibility is first increased and then decreased as we lower the chemical potential.
\begin{figure}[t]
\begin{center}
\includegraphics[scale=0.5, bb=200 180 400 380]{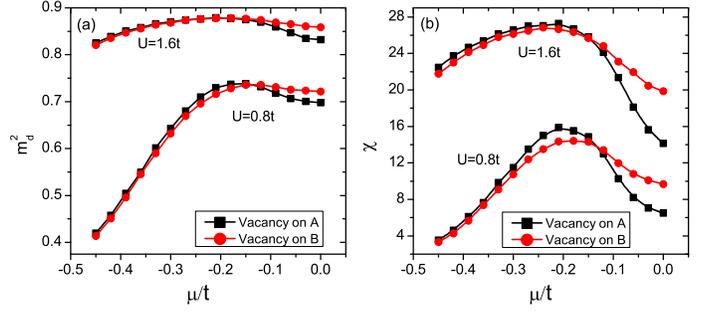}
\end{center}
\caption{(Color online). (a)The local moment and (b)the susceptibility for the two cases that vacancy on sublattice A and on sublattice B as a function of $\mu$. The impurity is located on the nearest site of the vacancy. In all the plates we use $V=0.5t$, $\varepsilon_{d}=-U/2$ and $\beta=1/T=40t^{-1}$.} \label{Fig:vacV0.5diffU}
\end{figure}
Presented in Fig. \ref{Fig:vacV0.5diffU} are the values of $m_{d}^{2}$ and $\chi$ for $V=0.5t$, $\beta=40t^{-1}$. The values of the local moment shown in Fig. \ref{Fig:vacV0.5diffU}(a) and the susceptibility shown in Fig. \ref{Fig:vacV0.5diffU}(b) are much larger than those for $V=1.0t$ case as shown in Fig. \ref{Fig:vacR0Layer1basicr}. However, the general behavior of the local moment and the spin susceptibility with respect to $\mu$ is the same as the case of $V=1.0t$ that for both A and B vacancy, the $m_{d}^{2}$ and $\chi$ are first enhanced and then suppressed as the chemical potential is lowered.
\begin{figure}[t]
\begin{center}
\includegraphics[scale=0.5, bb=180 180 400 430]{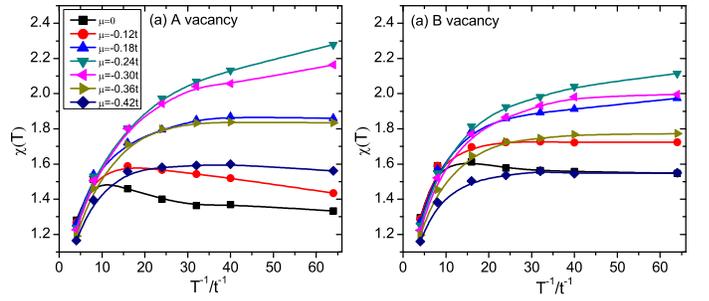}
\end{center}
\caption{(Color online). The spin susceptibility as a function of $T^{-1}$ for the case that (a) vacancy on sublattice A and (b) on sublattice B. The impurity is located on the nearest site of the vacancy. In all the plates we use $V=1.0t$, $\varepsilon_{d}=-U/2$.} \label{Fig:vackaiT}
\end{figure}
Fig. \ref{Fig:vackaiT} shows the spin susceptibility of the adatom as a function of inverse temperature for $V=1.0t$ and $\varepsilon_d=-U/2$. For both cases, we see that at low temperature, the susceptibility is temperature-independent at $\mu=0$. As $\mu$ is lowered, we can see that the values of $\chi$ tend to regain the features of Curie-like dependence. The reason is that as $\mu$ is moved away from the zero-energy, the LDOS of the carbon sites near the vacancy drops significantly and as a result, the screening of the local moment is suppressed. Hence, a well-developed local moment shows up on the impurity site.
However, if we go on lowering $\mu$, a temperature independent behavior would be regained. As we further lower the chemical potential, charges as well as the spins move out from the impurity and carbon sites, leading to a decrease in the spin susceptibility.

\begin{figure}[t]
\begin{center}
\includegraphics[scale=0.5, bb=200 190 400 420]{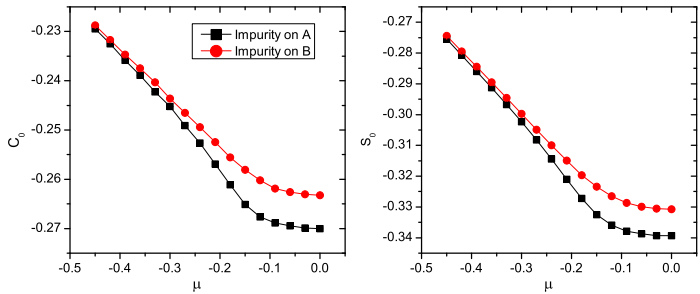}
\end{center}
\caption{(Color online). (a) The charge-charge correlation $C_0$ and (b) the spin-spin correlation $S_0$ with respect to $\mu$. For both cases the magnetic adatom is attached to the nearest site of the vacancy. We use $V=1.0t$, $U=0.8t$, $\varepsilon_{d}=-U/2$ and $\beta=1/T=40t^{-1}$.} \label{Fig:vacR0Layer1corr}
\end{figure}
In Fig. \ref{Fig:vacR0Layer1corr}, we plot the behavior of on-site correlation functions $C_0$ and $S_0$ with $\mu$ moving below the zero-energy. Here $C_0$ and $S_0$ is the correlation between the magnetic impurity and the carbon site where the impurity sits. We use $V=1.0t$, $U=0.8t$, $\varepsilon_{d}=-U/2$ and $\beta=1/T=40t^{-1}$. We can see that both the spin-spin and charge-charge correlations decrease in values as the chemical potential is lowered, as is shown in Fig. \ref{Fig:vacR0Layer1corr}. Both the correlations are the strongest at half-filling, and as the chemical potential is lowered, the correlations are reduced in amplitudes. As we have analyzed, vacancy induces zero-energy localized states on the surrounding sites and hence, at $\mu=0$, the impurity spin is strongly screened by the conduction electrons in the host material. As the chemical potential is lowered, both the impurity site and the carbon sites would lose the charge such that the charge-charge correlation and spin-spin correlation would decrease in values.
We can also find that both the spin-spin and charge-charge correlations between the magnetic impurity and the carbon site show larger amplitude for the A vacancy case. This is consistent with the results of the basic thermodynamic quantities presented in Fig. \ref{Fig:vacR0Layer1basicr}.

\section{Summary}

In summary, we studied an on-site magnetic impurity placed in the vicinity of a vacancy in the bilayer graphene with Bernal stacking. We mainly test two different cases, namely, vacancy located on A and B sublattices. We find that if the magnetic adatom is located on the adjacent site of the vacancy, the local moment is strongly suppressed due to the zero-energy mode which is induced by the vacancy. The local moment for the A vacancy case are slightly smaller than that in B vacancy case at $\mu \sim 0$, due to the broken symmetry caused by the Bernal stacking. As $\mu$ is lowered, both the spin and charge correlations between the adatom and carbon sites are weakened.
This behavior is completely different from that in pure BLG, where both of the correlations are enhanced as the chemical potential is lowered.

\section{Acknowledgement}
This work was supported by the Research Grants Council of Hong Kong (402310, HKUST3/CRF/09). F. M. Hu was supported by Academy of Finland through its Center of Excellence (2012-2017) program. We acknowledge the CPU time from CUHK in Hong Kong and CSC-IT Center for Science Ltd in Finland.

\bibliographystyle{phaip}
\bibliography{articles}

\begin{thebibliography}{10}

\bibitem{Castro10}
E.~V. Castro, M.~P. L\'opez-Sancho, and M.~A.~H. Vozmediano,
\newblock Phys. Rev. Lett. {\bf 104}, 036802 (2010).

\bibitem{Pereira06}
V.~M. Pereira, F.~Guinea, J.~M.~B. Lopes~dos Santos, N.~M.~R. Peres, and A.~H.
  Castro~Neto,
\newblock Phys. Rev. Lett. {\bf 96}, 036801 (2006).

\bibitem{Ugeda10}
M.~M. Ugeda, I.~Brihuega, and F.~J. M. G.-R. Guinea,
\newblock Phys. Rev. Lett. {\bf 104}, 096804 (2010).

\bibitem{Harigaya01}
K.~Harigaya,
\newblock Journal of Physics: Condensed Matter {\bf 13}, 1295 (2001).

\bibitem{Lehtinen04}
P.~O. Lehtinen, A.~S. Foster, Y.~Ma, A.~V. Krasheninnikov, and R.~M. Nieminen,
\newblock Phys. Rev. Lett. {\bf 93}, 187202 (2004).

\bibitem{Vozmediano05}
M.~A.~H. Vozmediano, M.~P. L\'opez-Sancho, T.~Stauber, and F.~Guinea,
\newblock Phys. Rev. B {\bf 72}, 155121 (2005).

\bibitem{Esquinazi03}
P.~Esquinazi et~al.,
\newblock Phys. Rev. Lett. {\bf 91}, 227201 (2003).

\bibitem{Ohldag07}
H.~Ohldag et~al.,
\newblock Phys. Rev. Lett. {\bf 98}, 187204 (2007).

\bibitem{Krasheninnikov07}
A.~Krasheninnikov and F.~Banhart,
\newblock Nature materials {\bf 6}, 723 (2007).

\bibitem{Gomez05}
C.~Gomez-Navarro et~al.,
\newblock Nature Materials {\bf 4}, 534 (2005).

\bibitem{hu2012}
F.~M. Hu, J.~E. Gubernatis, H.-Q. Lin, Y.-C. Li, and R.~M. Nieminen,
\newblock Phys. Rev. B {\bf 85}, 115442 (2012).

\bibitem{Withoff90}
D.~Withoff and E.~Fradkin,
\newblock Phys. Rev. Lett. {\bf 64}, 1835 (1990).

\bibitem{Hu11}
F.~M. Hu, T.~Ma, H.-Q. Lin, and J.~E. Gubernatis,
\newblock Phys. Rev. B {\bf 84}, 075414 (2011).

\bibitem{Hirsch86}
J.~E. Hirsch and R.~M. Fye,
\newblock Phys. Rev. Lett. {\bf 56}, 2521 (1986).

\bibitem{Rmp}
A.~H. Castro~Neto, F.~Guinea, N.~M.~R. Peres, K.~S. Novoselov, and A.~K. Geim,
\newblock Rev. Mod. Phys. {\bf 81}, 109 (2009).

\bibitem{Brandt88}
S.~M.~C. Brandt N.~B. and Y.~G. Ponomarev,
\newblock {\em Modern Problems in Condensed Matter Sciences},
\newblock North Holland, Amsterdam, 1988.

\bibitem{Dresselhaus02}
M.~S. Dresselhaus and G.~Dresselhaus,
\newblock Advances in Physics {\bf 51}, 1 (2002).

\bibitem{Choi08}
S.~Choi, B.~W. Jeong, S.~Kim, and G.~Kim,
\newblock Journal of Physics Condensed Matter {\bf 20} (2008).

\end{thebibliography}

\bibliographystyle{unsrt}  
\bibliographystyle{phaip}
\clearpage
\appendix

\end{document}